\documentclass{aa}
\usepackage{graphics}

\begin{document}

\thesaurus{12.03.4;12.03.3;12.04.01;12.12.1;11.03.1} 

\title{What Does Cluster Redshift Evolution Reveal ?
}
\author{A. Blanchard \and J.G. Bartlett}

\institute{ Observatoire Astronomique, 
            11, rue de l'Universit\'e, 67000 Strasbourg, France 
          }

\offprints{A.Blanchard}
\mail{blanchard@astro.u-strasbg.fr}
   
\date{Received \rule{2.0cm}{0.01cm} ; accepted \rule{2.0cm}{0.01cm} }

\titlerunning{Cluster Redshift Evolution}
\authorrunning{Blanchard \&  Bartlett}
\maketitle
 
\begin{abstract} 
Evolution of the cluster population has
been recognized as a powerful cosmological tool.  While 
the pre\-sent--day abundance of X-ray clusters
is degenerate in $\sigma_8$, $n$ and $\Omega_0$, 
Oukbir and Blanchard (1992, 1997) have pointed out that 
the number density evolution of X-ray clusters with redshift 
can be used to determine $\Omega_0$.  Here, we clarify 
the origin of this statement by identifying those
parameters to which the evolution of cluster number density 
is most sensitive.  We find that the evolution is controlled 
by only two parameters: the amplitude of fluctuations,
$\sigma_M$, on {\em the scale associated with the mass 
under consideration}, $R = 9.5h^{1/3} \Omega_0^{-1/3}$ 
$ M_{15}^{1/3}\; h^{-1}$Mpc, and the cosmological 
background density, $\Omega_0$.  In contrast, 
evolution is remarkably insensitive to the slope of 
the power spectrum.
We verify that the number density evolution of clusters 
is a powerful probe of the mean density of the universe,
under the condition that $\sigma_M$ is 
chosen to reproduce current-day abundances. 
Comparison of the cluster abundance 
at $z \sim 0.5-0.6$, from the EMSS, to the present-day 
abundance, from the ROSAT BCS sample, unambiguously reveals 
the existence of significant negative evolution. This number evolution,
in conjunction with the absence of any negative evolution  
in the luminosity-temperature relation, provides robust evidence 
in favor of  a critical density universe ($\Omega_0=1$), in agreement 
with the analysis by Sadat et al. (1998).
\end{abstract}  

\section {Introduction} 
 
X-ray galaxy clusters offer several interesting  ways to constrain 
cosmological parameters.  The temperature of the 
intra--cluster gas can be related to the virial mass according to
\begin{equation}
T = 4 M_{15}^{2/3}(1+z) {\rm keV}
\end{equation}
where $M_{15}\equiv M/10^{15} M_\odot$ and 
we hereafter assume $h = H_0/100.\mathrm{km s^{-1}Mpc^{-1}} = 0.5$. 
Such a relation can be easily deduced from 
the equation of hydrostatic equilibrium for the gas, 
leading to a temperature some 20\% larger 
than the value given in Eq. 1, which was inferred 
from numerical simulations (Evrard, Metzler \& Navarro 1996). 
Typical clusters have temperatures between 2 and 14 keV, 
corresponding to scales between 5 and 15 $h^{1/3}\Omega_0^{-1/3}
\; h^{-1}$Mpc.  Henry and Arnaud (1991) 
have shown how both the normalization and the slope of 
the power spectrum can be inferred from the local temperature 
distribution function, a technique which has been widely 
employed in recent years (see Bartlett 1997 for a review). 
Oukbir \& Blanchard (1992) proposed that the {\em evolution} of 
the X--ray cluster abundance is a powerful 
probe of the mean cosmological density, $\Omega_0$.  
In order to apply the technique, Oukbir \& Blanchard 
(1997, hereafter OB) established a detailed description of X-ray clusters.
This new approach based on evolution has also received much 
attention lately (Henry 1997; Bahcall et al. 1997); however, because 
the evolution of cluster 
number density depends in principle on the mass considered, 
the spectrum of the initial fluctuations, its normalization 
and the cosmological framework, doubts have been raised 
concerning the validity of the technique (Colafrancesco et al.
1997).  The purpose of this letter is to clearly identify the
parameters controlling cluster number density evolution 
and to examine what one may say about $\Omega_0$ 
using current data.

\section{Cluster properties and the mean density of the universe}

\begin{figure}
\resizebox{\hsize}{!}{\includegraphics{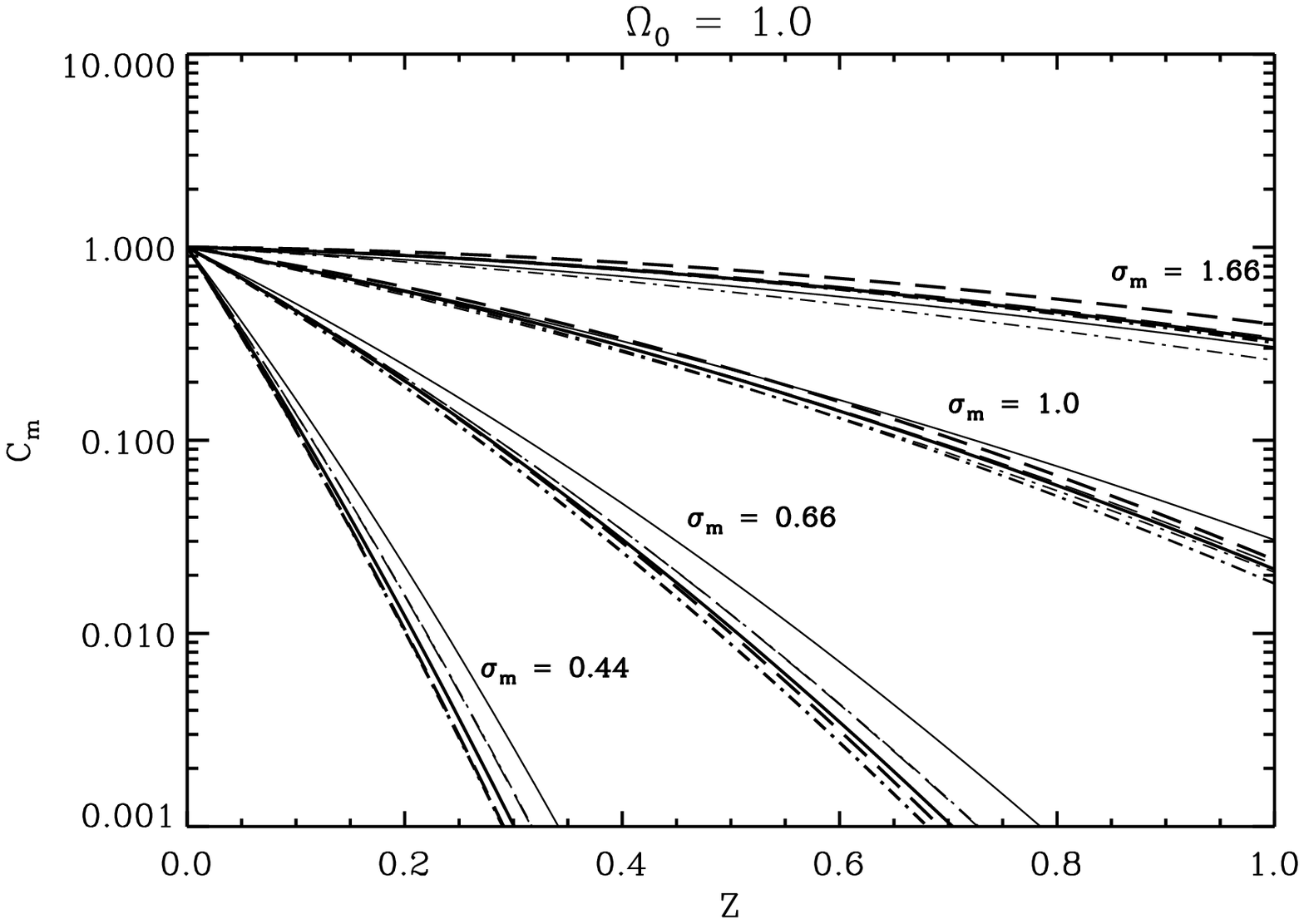}}
\resizebox{\hsize}{!}{\includegraphics{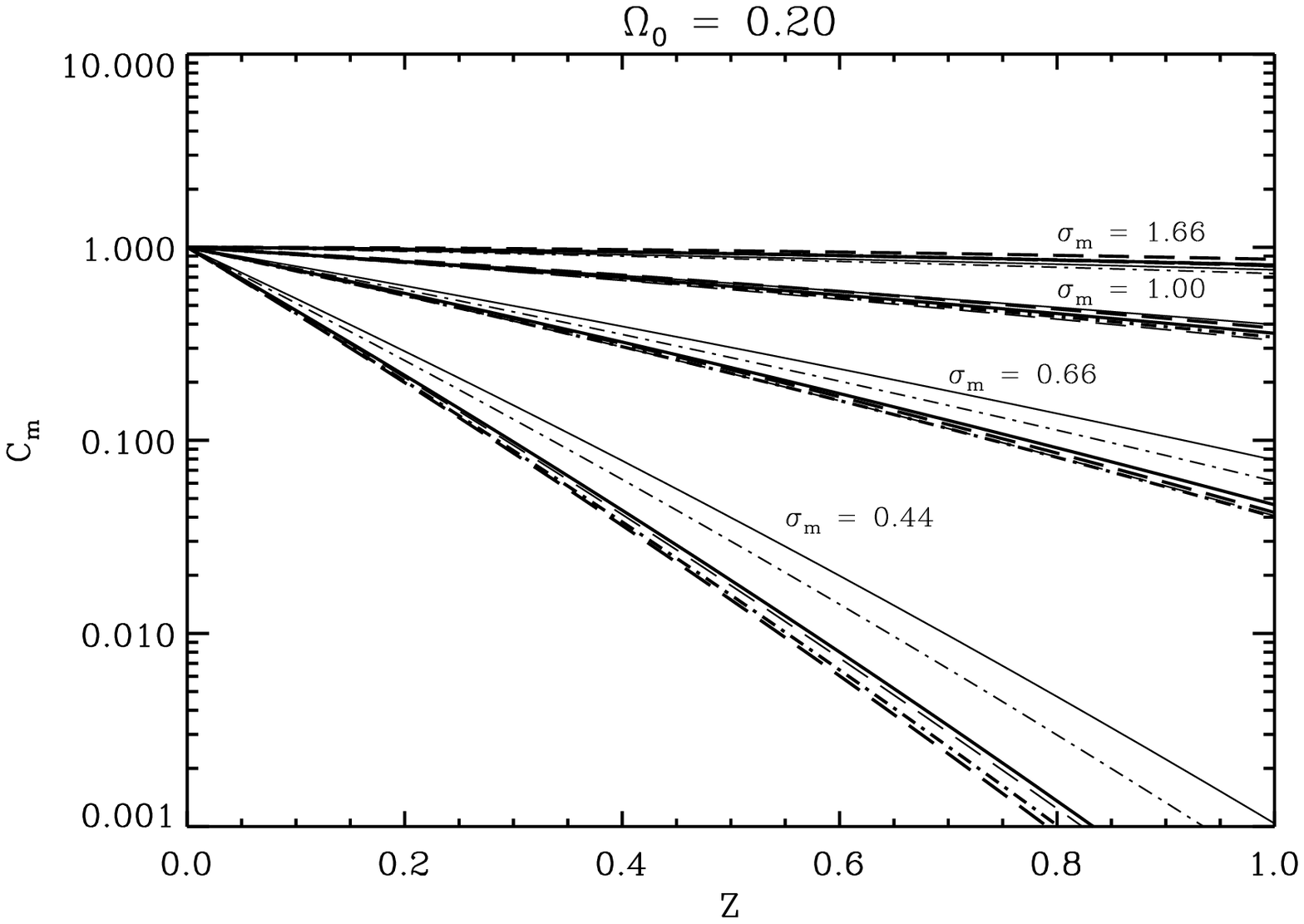}}
\caption{In this figure, we plot the evolution of the 
relative abundance of clusters, expressed 
by equation \protect\ref{eq:cm}, as a function of redshift.  
The continuous lines are for 
$M_{15} = 0.125$; the dashed lines, for $M_{15} = 1.0$; and 
the dot--dashed lines, for $M_{15} = 6.55$.  The thin 
lines correspond to an $n= 0$ power spectrum index, while the 
thick lines are for $n = -2$.  The upper 
graph represents $\Omega_0 = 1$, and the bottom is 
for $\Omega_0 = 0.2$.  In the first case, 
$\sigma_8$ varies from $0.35 \sigma_m$ to $2.56 \sigma_m$; in 
the second case, it changes from $0.82 \sigma_m$ to $5.5 
\sigma_m$.  Each figure illustrates that the relative 
abundance of clusters does not depend on the spectrum, 
nor on $\sigma_8$.  Clearly, the only important 
parameters are $\sigma_m$ and the cosmological density 
parameter, $\Omega_0 $, the latter evinced by the 
difference between the top and bottom panels.
}
\end{figure}
 
The Press-Schechter (1974) formalism, PS hereafter,  is a 
rather simple description of the mass function and its 
evolution, and it has been shown to be in good agreement 
with numerical simulations (Lacey \& Cole 1994). 
 The PS prescription is:
\begin{eqnarray}
\nonumber
N(M,z) dM & = &
- \sqrt{{2\over \pi}}{{\rho_c}\Omega_0 \over M^2}
{\delta_{c}A\over \sigma_M}
 {d\ln\sigma\over d\ln M}           
 \exp \left(-{(\delta_{c}A)^2\over 
2\sigma^2_M}\right) dM
\nonumber
\end{eqnarray}
where $A = A(\Omega_0,z)$ is the growth rate of linear 
density perturbations, $\rho_c$ is the Einstein--de Sitter
density, $\delta_c = \delta_c(\Omega_0,z)$ is the critical linear 
over--density
required for collapse and $\sigma_M$ is the present-day amplitude of 
density perturbations on a scale $M$.  
From this expression, it is 
clear that the cluster abundance at redshift $z= 0$, 
for a given mass $M$, determines $\sigma_M$ 
almost independently of the value of the spectral index, 
$n$, or of the density parameter.
There is only a slight dependence on these quantities
due to their presence in the pre--factor of 
the exponential term (and there is almost no influence from
a possible cosmological constant).  In practice, matching 
the present-day number of observed clusters with $T \geq  
4$ keV requires $\sigma(10^{15}M_{\odot})  \sim 0.6$ in 
an $\Omega_0 = 1$ universe, and a similar value,
$\sigma (10^{15}M_{\odot})  \sim 0.8$, in an 
$\Omega_0 = 0.2$ universe.  However, it should be kept in 
mind that this corresponds to two different linear scales 
of the initial density perturbation field:
\begin{equation}
R \approx 9.5 h^{1/3}\Omega_0^{-1/3}
 M_{15}^{1/3}\; h^{-1}\mathrm{Mpc} 
\end{equation}
the difference being almost a factor of 2 between 
$\Omega_0 = 1$ and $\Omega_0 = 0.2$.  This means that 
the abundance of $10^{15}{M_{\odot}}$ clusters determines
the amplitude on {\em different} linear scales.  In an 
$\Omega_0 = 1$ cosmology, $\sigma$ is fixed on a scale 
of $8h^{-1}\mathrm{Mpc}$, the traditional normalization 
scale, while in an $\Omega_0 = 0.2$ cosmology, 
$\sigma$ is instead set on a scale of $15h^{-1}\mathrm{Mpc}$.
Accordingly, for an open model ($\Omega_0 \sim0.2$),
$\sigma(8h^{-1}\mathrm{Mpc})$ must be found by 
extrapolation using a specific $n$, and is uncertain by a 
factor of two (see OB, Fig. 1).

	Let's now  examine what governs the  redshift 
evolution of the cluster abundance on a given mass scale, 
$M$.  As inspection of Eq. 1 clearly shows, 
only $\sigma_M$, {\em on the mass--scale 
considered}, $A$ and $\delta_c$ govern 
the evolution with redshift.  As the the later
two quantities only depend on $\Omega$, the redshift 
evolution is completely independent of the power spectrum 
index, $n$, and does not depend explicitly on 
the normalization at 8h$^{-1}$Mpc.  This makes 
the redshift evolution remarkably simple to understand and 
to employ as a cosmological probe:  once $\sigma(M)$
is set by the present-day cluster abundance, the evolution of 
the number of clusters on the same mass scale is entirely and 
uniquely determined by the cosmological background ($\Omega_0$).  
This is the essential reason for the robustness of the 
cosmological test originally proposed by Oukbir and 
Blanchard (1992).  In order to illustrate this point, 
we define the quantity  
\begin{equation}
C_m(z) = \frac{n(>M,z)}{n(>M, z=0.)} 
\label{eq:cm}
\end{equation}
as a simple measure of redshift evolution.  We plot
$C_m(z)$ in Fig. 1 for different spectra normalized 
to the same amplitude, $\sigma(M)$, and for 
two different cosmological background 
densities - $\Omega_0 = 1$ and  $\Omega_0 = 0.2$.
It is important to note that, for fixed $\sigma(M)$, 
$\sigma_8$ varies as $n$ and $M$ change; for example,  
when $\sigma_M = 1$, the normalization $\sigma_8$ goes from 
0.3 to 1.3 as $n$ is varied from $n=-2$ to $n=-1$ for the
range of different masses mentioned in the figure.
In other words, the `bundle' of curves corresponding to 
each value of $\sigma(M)$ covers a large range of
$M$, $n$ and $\sigma_8$.  The fact that
the curves fall into tight bundles defined 
only by $\sigma(M)$ confirms what we have inferred from Eq. 1:  
for a given value of $\Omega_0$, the redshift evolution is almost 
completely independent of $n$ and and does not depend directly 
$\sigma_8$.  On the other 
hand, there is a significant difference between the two 
cosmological models -- as much between the open and critical 
models shown as between  $\sigma_m = 0.66$ and $\sigma_m = 1.0$ -- a 
difference significant enough to potentially discriminate 
between the two cosmologies. 

\section{Comparison with observations}

	In order to apply this technique, it is 
important to notice that the mass in the PS formula 
corresponds to a {\em fixed contrast density} and, therefore,
represents very different objects at different 
redshifts.  As this mass is not directly observable, 
we must resort to some other, more observable cluster quantity.
To this end, we use cluster 
temperature and introduce an evolution coefficient: 
\begin{equation}
c_T(z)= \frac{n(>T,z)}{n(>T, z=0.)} 
\end{equation}
which we will consider for two {\em temperatures} - 4 and 6 keV.  
One must then
take into account the fact that clusters with identical 
temperatures at different redshifts 
correspond to different masses (in the PS language - see 
Eq. 1); thus, the
evolution expressed in terms of temperature 
could in principle be sensitive to the spectrum.  

	We estimate our modeling uncertainty using the results of
Oukbir et al. (1997, hereafter OBB) and OB.  For $\Omega_0 = 1$, we allow 
 $\sigma(M=10^{15}{\rm M_{\odot}})$ to cover the range 
 $0.55 -0.65 $, i.e. $\sigma_8$ in the range  $ 0.53 -0.625 $ 
in agreement with Viana and Liddle (1996).  
Rather than the best fitting value of
$n = -1.8$ given by OBB, we use $n=-1.4$, because it
is closer to a $\Gamma$--CDM model with $\Gamma= 0.25$;
this reduces the amount of evolution by a factor of 2
at $z \sim 0.5$, relative to the $n=-1.8$ case.
For the open model, we set  
 $\Omega_0 = 0.3$ and consider two extreme cases with 
$\sigma(M=10^{15}{\rm M_{\odot}}) = 0.78$ (according to
the results of OB) : $n= -1.8$ with $\sigma_8 
= 1.07 $, and $n= -1.2$ with $\sigma_8 = 0.94$. 
For these ranges of parameters, we examine the redshift 
evolution of the cluster number density 
for 4 keV and 6 keV  clusters. 
The corresponding range of  
predictions for $c_T$ 
are presented as the grey areas in Fig. 2.
Notice that $\sigma(10^{15}{\rm M_{\odot}})$ 
differs slightly between the two models (see the previous 
section), increasing the evolutionary difference between them.
As one can see, the uncertainty for $\Omega_0 = 1$ is rather large,
but the probe can certainly discriminate between 
a low--density and a high--density universe. \\
	It is difficult to directly apply this test to 
present--day X--ray cluster samples, because this
requires knowledge of the temperature distribution function 
at high $z$.  The only well controlled 
high--redshift sample of X--ray clusters is the EMSS 
(Gioia \& Luppino 1994).  It has been studied and modeled 
in detail by OB.  They concluded that, in order {\em to 
self--consistently model X--ray clusters in an open 
universe, one must introduce negative evolution 
in the temperature--luminosity relation} (i.e., at a given
temperature, clusters are less luminous in the past).  
The reason is that the EMSS sample provides definitive 
evidence for negative evolution of the X-ray 
luminosity function (see the following discussion), while 
an open cosmological model would predict an X--ray temperature
function with little evolution.  

	Recently, several authors have quoted numbers for the
redshift evolution of the cluster number density.  
Carlberg et al. (1997) have estimated the 
number density of CNOC clusters with velocity dispersions 
$\geq 800 $ km/s.  They find $n({\overline z} = 0.22) = 4.38\times 10^{-8.}$
and $n({\overline z} = 0.45) = 1.13\times 10^{-8.}$. 
We may convert the velocity dispersion to an X-ray 
temperature of $T_x \approx 5.$ keV (in agreement with their 
luminosity of $L_{[0.3-3.5]} \sim 4. 10^{44}$ erg/s)
using the conversion provided by Sadat et al. (1998), which
shows good agreement with recent ASCA measurements (although
a few clusters appear discrepant). 
Henry (1997) provides the first actual estimate of evolution of 
the temperature distribution function, although at moderate redshift 
($z \approx 0.35$); the data seem to indicate a significant 
amount of evolution.  Fan et al. (1997), using the CNOC sample, find 
\begin{equation} 
\frac{n( z=0.5)}{n(z=0.)} \approx 0.2
\end{equation}
for clusters of mass $M_{1.5}= 6.3 10^{14}\rm 
h^{-1}M_{\odot}$ {\em within a physical radius 
of $1.5 \rm h^{-1}Mpc$}.  This corresponds to an 
approximate temperature of 4.5 keV for a virialized cluster
(independent of redshift).  

\begin{figure}
\resizebox{\hsize}{!}{\includegraphics{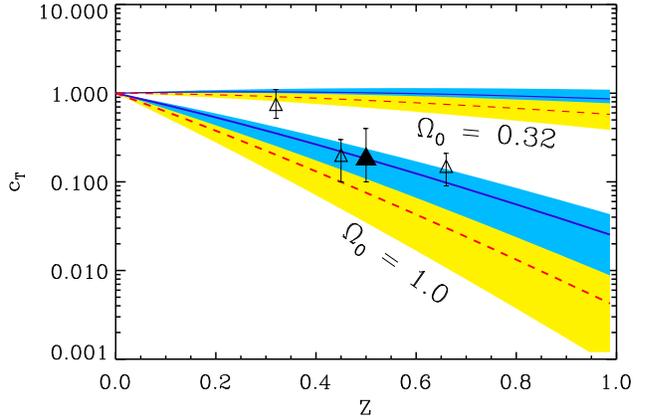}}
\caption{Relative evolution of the abundance of clusters
above a given temperature.  The continuous and dashed lines 
show the cluster abundance evolution for $ T > 4$ keV
and $ T > 6$ keV, respectively  The grey 
area delimits our estimate of the modeling uncertainty, taken
from Oukbir \& Blanchard (1997) and  Oukbir et al.  
(1997).  The triangles are from the observations as given 
by Carlberg  al. (1997) and Bahcall 
et al. (1997).  The point at $z = 0.66$ is derived from 
the luminosity function of the EMSS. The open  triangles 
correspond to clusters with $T > 6$ keV, while the filled triangle is
for $T > 4$ keV, assuming no evolution in the $L_x-T_x$ relation. }
\end{figure}
	The abundance of X-ray clusters at redshift 0.66 
can be estimated from the EMSS (Luppino and Gioia 1995):
\begin{equation} 
N(L_{[0.3-3.5]}>5. 10^{44}) \approx 1. 10^{-8}
\end{equation}
In the absence of evolution in the $L_x-T_x$ 
relation , such clusters would have temperatures greater 
than $5$ keV (Arnaud and Evrard 1997). The abundance of clusters 
deduced from the temperature distribution function at $z = 0$
is, rather surprisingly, highly uncertain (see, for instance, Table 1 in 
Carlberg et al., 1997). To lower this uncertainty, we estimate 
the present--day abundance of similar clusters from the 
BCS {\em luminosity function}, which is constructed 
from a much larger cluster sample, (Ebeling et al., 1997).  
In the ROSAT band - $[0.1,2.4]\;$ keV - 4 to 6 keV clusters have 
a luminosity greater than $4. 10^{44}  {\rm erg/s/cm^2}$,
yielding $N(>L) \sim 0.6 \; 10^{-7}$, giving: 
\begin{equation}
\frac{n_T( z=0.66)}{n_T( z=0.00)} \approx 0.16 \pm0.06
\end{equation}
which is direct and clean  evidence for some kind of evolution. 
 The above density  will serve as our
reference for the abundance at $z = 0$ for the  CNOC clusters: 
$c_T( {\overline z}=0.27)  \approx 0.44^{+0.29}_{-0.15}$
and 
$c_T( {\overline z}=0.45) \approx 0.11 ^{+0.075}_{-0.045}$. 

	As we have already mentioned, open models ($\Omega \sim 0.2$), 
for which the temperature distribution function shows little
evolution, {\em cannot} be consistent with the EMSS 
distribution unless there is strong negative 
evolution of the luminosity-temperature relation: 
whatever the value of $\Omega_0$, the properties of the 
cluster population (either the number density or the 
luminosity--temperature relation) must evolve in order to explain the 
EMSS redshift distribution.
One may wonder 
whether a bias in the EMSS sample could lead to a severe underestimation
of the cluster abundance at large $z$.  This seems rather unlikely,
for at such redshifts clusters are almost point--like compared 
to the size of the detection cell (5'); furthermore, no 
systematic bias has been found in the photometry
(Nichol et al., 1997). 

	These numbers already give interesting insight concerning 
the density parameter of the universe.  It is clear from Fig. 2 
that the critical model is favored over a low--density model, 
according to the cluster abundances reported in the 
recent literature; however, a note of caution: it must be remembered 
that in all cases, the data were analyzed assuming, either 
implicitly or explicitly, a non--evolving relation between 
temperature and luminosity.  It is for this same reason that our
present conclusions are exactly the same as those
given by Oukbir and Blanchard (1997): under the assumption 
of a non-evolving temperature-luminosity relation, the EMSS 
redshift distribution of X-ray clusters favors a high density 
universe.  This result is supported by the additional
information that available data 
on distant X-ray clusters does not demonstrate any  
sign of the strong negative evolution of the luminosity--temperature
relation needed to save the open model (Sadat et al., 1998)
(this is independent of the possible addition of a cosmological constant).
This additional piece of information is critical to the
conclusion, because without it, we have no way of 
understanding the flux limited selection of the EMSS in
terms of temperature.

\section{Conclusion and discussion} 

	The purpose of this letter was to clarify
the nature of the evolution of the cluster 
temperature distribution function.  As we have seen, this 
evolution depends primarly on the 
amplitude of the fluctuations  on the scale 
under consideration, $\sigma(M)$, and the cosmological 
background, $\Omega_0$. This is the origin of the robustness of 
the cosmological test initially proposed by Oukbir and 
Blanchard (1992).  
The EMSS redshift 
distribution, as modeled by 
Oukbir and Blanchard (1997),  combined with the absence of observed 
negative evolution in the 
temperature--luminosity relation 
provided  the first evidence 
for a high density universe from this technique (Sadat et al 1998).  Our analysis leads to a 
similarly high value for the density of the universe.
During the submission of this letter,  we learned that similar 
conclusions were reached
by two other groups who included ROSAT cluster redshift distributions
(Borgani et al., 1998; Reichart et al, 1998). 
Because this test is primary sensitive to the dynamical
behavior of the universe as a whole (through the growth 
rate of linear density fluctuations), we consider this to be
the strongest evidence in favor of a critical density universe
presently available.
\begin{acknowledgements}
We aknowledge R.~Sadat for useful comments. One of us, A.B., aknowledge the 
hospitality of the CAUP, Porto, where this work was finalized.\vspace*{-4mm} 
\end{acknowledgements}

\end{document}